\pdfoutput=1
\PassOptionsToPackage{backref,pagebackref}{hyperref}
\documentclass[11pt]{article}

\usepackage[top=30truemm,bottom=30truemm,left=25truemm,right=25truemm]{geometry}

\usepackage[utf8]{inputenc}
\usepackage[T1]{fontenc}
\usepackage{textcomp}

\usepackage{amsmath}
\usepackage{amssymb}
\usepackage{amsthm}
\usepackage{amsfonts}
\usepackage{mathtools}

\usepackage{booktabs}
\usepackage{flafter}
\usepackage{enumitem}

\usepackage[authoryear,round]{natbib}
\makeatletter
\@ifundefined{NAT@citetp}{}{%
  \let\orig@NAT@citetp\NAT@citetp
  \def\NAT@citetp{\def\NAT@spacechar{\nobreakspace}\orig@NAT@citetp}}
\makeatother

\usepackage[svgnames]{xcolor}
\usepackage{tikz}
\usetikzlibrary{decorations.pathreplacing}
\usepackage[hypertexnames=false,backref,pagebackref]{hyperref}
\hypersetup{
  colorlinks=true,
  linkcolor=blue,
  citecolor=blue,
}

\IfFileExists{tgtermes.sty}{\usepackage{tgtermes}}{}

\usepackage[ruled,vlined,linesnumbered]{algorithm2e}
\SetArgSty{textnormal}

\usepackage{microtype}
\usepackage[labelfont=bf]{caption}
\usepackage{url}

\usepackage[nameinlink,capitalise,noabbrev]{cleveref}
\crefname{equation}{}{}
\Crefname{algocf}{Algorithm}{Algorithms}
\crefname{appendix}{Appendix}{Appendices}
\Crefname{appendix}{Appendix}{Appendices}
\crefname{line}{Line}{Lines}
\Crefname{line}{Line}{Lines}
\newcommand{\eqpart}[2]{\hyperref[#1]{(\ref*{#1}#2)}}

\theoremstyle{plain}
\newtheorem{theorem}{Theorem}
\newtheorem{lemma}{Lemma}
\newtheorem{corollary}{Corollary}
\newtheorem{proposition}{Proposition}
\theoremstyle{definition}
\newtheorem{remark}{Remark}

\renewcommand{\tilde}{\widetilde}

\def\R{\mathbb{R}}
\DeclareMathOperator*{\argmax}{arg\,max}

\newcommand{\Reg}{\mathsf{Reg}}

\newcommand{\one}{\mathbf{1}}

\title{Optimal Regret Lower Bounds in Matrix Games}
\author{
  Taira Tsuchiya\footnote{
    The University of Tokyo, The University of Osaka, and RIKEN;
    \texttt{tsuchiya@mist.i.u-tokyo.ac.jp}.
  }
}
\date{}

\begin{document}
\maketitle

\begin{abstract}
We study learning in two-player zero-sum games, also known as matrix games, in which two players with $m$ and $n$ actions repeatedly select mixed strategies for $T$ rounds.
The learning dynamics based on optimistic Hedge guarantee regrets of $O(\log(mn))$, which improve to $O(\sqrt{\log m\log n})$ if each player knows the number of actions of the opponent, whereas matching regret lower bounds have been known only for specific learning dynamics.
We prove that, for every deterministic learning dynamic, there exists a payoff matrix for which the maximum individual regret is $\Omega(\sqrt{\log m\log n})$ for a wide range of $m$ and $n$.
Together with the upper bound $O(\sqrt{\log m\log n})$, this lower bound shows that the minimax regret of deterministic learning dynamics in matrix games is $\Theta(\sqrt{\log m\log n})$ in this regime.
For all $m$, $n$, and $T$, we also prove the lower bound $\Omega(\min\{T,\log\min\{m,n\}\})$.
These lower bounds hold even if each player also observes the strategies and the feedback of the opponent.
Our main technical contribution is an adaptive construction of the payoff matrix.
While running the algorithms of the two players, we add in each round a rank-one matrix that keeps the feedback of the previous rounds unchanged and increases a lower bound on the sum of the regrets of the two players.
As by-products, we obtain lower bounds on the rate at which the average strategies converge to a Nash equilibrium and on the query complexity of matrix-vector products that deterministic algorithms need to compute an approximate Nash equilibrium.
\end{abstract}

\section{Introduction}\label{sec:introduction}
We study learning in two-player zero-sum games, also known as matrix games.
In this problem, two players repeatedly play a game given by an unknown $m\times n$ payoff matrix.
In each of $T$ rounds, the two players select mixed strategies over $m$ and $n$ actions, respectively, and each player then observes the expected payoff of each of its actions against the strategy of the opponent as feedback.
Using this feedback, each player updates its strategy with the aim of minimizing its regret.
If both players use no-regret learning algorithms, the average strategies of the two players converge to a Nash equilibrium at a rate determined by their regrets~\citep{freund99adaptive,cesabianchi06prediction}.

A standard no-regret algorithm is Hedge~\citep{littlestone94weighted,freund97decision}, which guarantees a regret of $O(\sqrt{T\log m})$ for the player with $m$ actions against an arbitrary opponent.
This worst-case bound can be significantly improved if both players use suitable learning algorithms.
\citet{daskalakis11near} give the first no-regret learning algorithm whose regret is logarithmic in $T$ and polylogarithmic in $m$ and $n$ when both players use it.
Subsequently, the learning dynamic based on optimistic Hedge~\citep{rakhlin13optimization,syrgkanis15fast} with constant learning rates was shown to guarantee that the regrets of both players are $O(\log(mn))$, which is independent of $T$.
This implies that the average strategies converge to a Nash equilibrium at the rate $O(\log(mn)/T)$.
If each player additionally knows the number of actions of the opponent, these regrets can be further improved to $O(\sqrt{\log m\log n})$~\citep{tsuchiya25tight}.

In contrast to these upper bounds, regret lower bounds in matrix games are far less understood.
Most existing regret lower bounds are specific to particular learning algorithms such as Hedge~\citep{syrgkanis15fast,chen20hedging}.
For the learning dynamic based on optimistic Hedge, the dependence on the numbers of actions $m$ and $n$ in the above upper bounds is known to be tight~\citep{tsuchiya25tight}, but this does not exclude the possibility that other learning dynamics achieve smaller regrets.
For strongly-uncoupled learning dynamics, \citet{daskalakis11near} show a lower bound of $\Omega(1/T)$ on the rate at which the average strategies converge to a Nash equilibrium, but this lower bound does not depend on $m$ and $n$.
This raises a natural question:
\begin{center}
\emph{What is the minimax optimal regret in matrix games?}
\end{center}

\paragraph{Contributions of this paper.}
We answer this question for deterministic learning dynamics by proving regret lower bounds that match the above upper bounds.
Our lower bounds hold even if each player also observes the strategies and the feedback of the opponent in all previous rounds.

We first show that, for every deterministic learning dynamic, there exists a payoff matrix for which the maximum individual regret is $\Omega(\min\{T,\log\min\{m,n\}\})$~(\cref{cor:logarithmic}).
Together with the upper bound of optimistic Hedge, this lower bound shows that the minimax value of the maximum individual regret for deterministic learning dynamics is $\Theta(\min\{T,\log m\})$ when the numbers of actions are the same, that is, $m=n$.
Based on the construction of this lower bound, we further strengthen it to $\Omega(\sqrt{\log m\log n})$ when $m\ge n$, $\log m\le n^\alpha$ for a constant $\alpha\in(0,1)$, and $T\ge\log_2m$~(\cref{cor:rectangular}).
This lower bound matches the upper bound $O(\sqrt{\log m\log n})$ for players who know the number of actions of the opponent~\citep{tsuchiya25tight}.
The condition $\log m\le n^\alpha$ cannot be removed entirely, since, for $n=2$, there is a deterministic learning dynamic whose regrets are bounded by a constant for every $m$~(\cref{prop:two-columns} in~\cref{app:two-columns}).

Our main technical contribution is the construction of the payoff matrices in these lower bounds.
Since the strategies selected by a learning dynamic depend on the payoff matrix through the feedback, we construct the matrix adaptively while running the algorithms of the two players for the first $k$ rounds.
At each round $t=1,\ldots,k$, after the algorithms select their strategies $x_t$ and $y_t$, we define
\begin{equation}
\label{eq:At_intro}
A_t=A_{t-1}+\frac13(u_t-u_{t-1})v_t^\top,
\end{equation}
where $A_0=0$ is the zero matrix, $u_0$ is the all-one vector, and $u_t\in[-1,1]^m$ and $v_t\in[-1,1]^n$ are vectors orthogonal to $x_1,\ldots,x_t$ and to $y_1,\ldots,y_t$, respectively.
The algorithms receive the feedback of round $t$ based on the matrix $A_t$, and the payoff matrix is $A=A_k$.
By the orthogonality, the terms added in later rounds do not change this feedback, and hence the deterministic algorithms select the same strategies when they are run on $A$.
We choose the vectors so that the entries of $A$ are in $[-1,1]$ and so that, for some actions $i_*$ and $j_*$, the coordinates $u_t(i_*)$ and $v_t(j_*)$ are equal to $1$ until round $k$.
Then, on the matrix $A$, the action $i_*$ has payoff zero against every strategy of the $y$-player, whereas each strategy $x_t$ has payoff $-1/3$ against the action $j_*$.
This construction allows us to lower bound the sum of the regrets of the two players after $H$ rounds by $H/3$ for every $H\le k$~(\cref{lem:matrix} in~\cref{sec:construction}).

The number of rounds $k$ is limited by the requirement that the coordinates $u_t(i_*)$ and $v_t(j_*)$ are equal to $1$ in all rounds.
Under the orthogonality constraints, only about half of the coordinates that are still equal to $1$ can be kept equal to $1$ in each round, and hence $k$ can be at most about $\log_2\min\{m,n\}$~(\cref{lem:nested} in~\cref{sec:proof-basic}).
This is the reason why our lower bounds are logarithmic in the numbers of actions.
For the lower bound $\Omega(\sqrt{\log m\log n})$ in~\cref{cor:rectangular}, in which $n$ can be polylogarithmic in $m$, we avoid this halving for the $y$-player.
In place of the action $j_*$, we use a mixed strategy of the $y$-player that is updated multiplicatively in each round, at the cost of a smaller increase of the lower bound per round~(\cref{lem:rectangular-comparator} in~\cref{sec:proof-rectangular}).

Our lower bounds also have consequences for the computation of Nash equilibria.
They give a lower bound of $\Omega(\log\min\{m,n\}/T)$ on the rate at which the average strategies of a deterministic learning dynamic converge to a Nash equilibrium~(\cref{cor:convergence}), which improves the lower bound of~\citet{daskalakis11near}.
When $m=n$, this lower bound matches the rate $O(\log(mn)/T)$ of optimistic Hedge up to a constant factor.
As a by-product of our lower bound, we also show that the query complexity of matrix-vector products that deterministic algorithms need to compute an $\varepsilon$-approximate Nash equilibrium is $\Omega(\log\min\{m,n\})$ when $\varepsilon\le1/12$~(\cref{cor:query}).
This can be slightly larger than the existing lower bounds by~\citet{hadiji23towards,kornowski26oracle}, and \Cref{fig:query} in~\cref{sec:discussion} compares this lower bound with the known upper and lower bounds.

\paragraph{Related work.}
On the algorithmic side, optimistic Hedge is an instance of optimistic follow-the-regularized-leader~\citep{rakhlin13optimization}, and the use of optimistic predictions has become a central approach to the design of learning dynamics to achieve fast convergence~\citep[\textit{e.g.,}][]{syrgkanis15fast,chen20hedging,daskalakis21near,anagnostides22uncoupled,farina22near,soleymani25cautious,tsuchiya26sublogarithmic}.
Under the feedback considered in this paper, standard optimistic Hedge and optimistic follow-the-regularized-leader dynamics with deterministic initialization and parameter choices generate their mixed strategies deterministically and are therefore covered by our lower bounds.

On the lower bound side, \citet{syrgkanis15fast} show that the regret of a player who uses Hedge with an arbitrary fixed learning rate against a best-responding opponent is $\Omega(\sqrt T)$ in some two-player zero-sum game.
\citet{chen20hedging} show that, when both players use Hedge with a common fixed learning rate, the regret of at least one of the players cannot be guaranteed to be $o(\sqrt T)$ in two-player general-sum games.
When both players use optimistic Hedge with learning rates $\eta$ and $\eta'$, their regrets can be at least $(\log m)/\eta-o(1)$ and $(\log n)/\eta'-o(1)$ in a two-player zero-sum game~\citep{tsuchiya25tight}.
\citet{anagnostides24complexity} prove conditional, dimension-dependent lower bounds for Hedge and optimistic Hedge in multiplayer general-sum games under the Exponential Time Hypothesis for PPAD.
These lower bounds are for specific learning algorithms, whereas our lower bounds are unconditional and hold for every deterministic learning dynamic.

A related problem is the query complexity of matrix games, that is, the number of matrix-vector products with the payoff matrix that are needed to compute an $\varepsilon$-approximate Nash equilibrium~\citep{nemirovski04prox,nesterov05smooth,hadiji23towards,kornowski26oracle,karmarkar26solving}.
This problem differs from regret minimization in that the output of the algorithm is evaluated.
A consequence of our regret lower bound for this problem is discussed in~\cref{sec:discussion}.

A payoff matrix as a sum of rank-one matrices of the form $(u_t-u_{t-1})v_t^\top$, as in~\cref{eq:At_intro}, is also used by~\citet{kornowski26oracle} to prove a lower bound on the query complexity of matrix-vector products that deterministic algorithms need to compute an approximate Nash equilibrium of a matrix game.
The two constructions share this form of the matrix, but the vectors in the rank-one terms are chosen in different ways and for different purposes.
\citet{kornowski26oracle} obtain these vectors by projecting Gaussian vectors onto the orthogonal complement of the previous queries, and show that the output of the algorithm after $N$ matrix-vector products is suboptimal by $\Omega(N^{-3/2})$ up to logarithmic factors.
Our matrices are instead constructed so that a lower bound on the sum of the regrets increases by a constant in every round of the construction, and the vectors $u_t$ and $v_t$ are chosen for this purpose as we will see in the following sections.

\section{Problem Setting and Preliminaries}\label{sec:setting}
This section introduces the problem setting and the relation between no-regret learning and Nash equilibria, and formalizes the class of learning dynamics considered in this paper.
For a positive integer $n$, we write $[n]=\{1,\ldots,n\}$ and
$\Delta_n=\{y\in\R^n:y(j)\ge0\text{ for all }j\in[n],\ \sum_{j=1}^n y(j)=1\}$.
We use $\one$ to denote the all-one vector and $e_i$ to denote the $i$-th standard basis vector, with their dimensions determined by the context.
For a vector $x$, we write $x(i)$ for its $i$-th coordinate, and $\|x\|_p$ for its $\ell_p$-norm.
All logarithms are natural unless the base is specified.

\paragraph{Setup.}
Learning in a two-player zero-sum game, also known as a matrix game, is characterized by a payoff matrix
$A\in[-1,1]^{m\times n}$, where $m$ and $n$ denote the number of actions of the $x$- and $y$-players, respectively.
At each round $t=1,\ldots,T$, the $x$-player selects a strategy $x_t\in\Delta_m$ and the $y$-player selects $y_t\in\Delta_n$ simultaneously.
Then, the $x$-player observes an expected gain vector $g_t=Ay_t$ and the $y$-player observes an expected loss vector $\ell_t=A^\top x_t$.
The $x$-player's payoff and the $y$-player's loss are both $\langle x_t,Ay_t\rangle$.

The goal of each player is to minimize the (external) regret.
For any $H\in[T]$, the regrets of the $x$- and $y$-players over the first $H$ rounds are given by
\[
\Reg_H^x
=\max_{x^*\in\Delta_m}\sum_{t=1}^H\langle x^*-x_t,Ay_t\rangle,
\qquad
\Reg_H^y
=\max_{y^*\in\Delta_n}\sum_{t=1}^H\langle y_t-y^*,A^\top x_t\rangle.
\]
The social regret is defined as the sum of the regrets of both players, that is,
\begin{equation}\label{eq:comparator}
\begin{split}
\Reg_H^x+\Reg_H^y
&=\max_{x^*\in\Delta_m,\;y^*\in\Delta_n}\sum_{t=1}^H\bigl(\langle x^*,Ay_t\rangle-\langle x_t,Ay^*\rangle\bigr)\\
&=H\biggl(\max_{x\in\Delta_m}\langle x,A\bar y_H\rangle-\min_{y\in\Delta_n}\langle\bar x_H,Ay\rangle\biggr),
\end{split}
\end{equation}
where $\bar x_H=\frac1H\sum_{t=1}^Hx_t$ and $\bar y_H=\frac1H\sum_{t=1}^Hy_t$ are the average strategies over $H$ rounds.
We will prove our lower bounds by fixing a pair $(x^*,y^*)$ in the maximum in~\cref{eq:comparator}.
Note that a lower bound on the social regret also gives a lower bound on the maximum individual regret, since $\max\{\Reg_H^x,\Reg_H^y\}\ge(\Reg_H^x+\Reg_H^y)/2$.

\paragraph{No-regret learning and Nash equilibrium.}
A pair of probability distributions $(x^*,y^*)$ over action sets $[m]$ and $[n]$ is said to be an \emph{$\varepsilon$-approximate Nash equilibrium} for $\varepsilon\ge0$ if for any distributions $x\in\Delta_m$ and $y\in\Delta_n$,
\[
\langle x,Ay^*\rangle-\varepsilon
\le\langle x^*,Ay^*\rangle
\le\langle x^*,Ay\rangle+\varepsilon.
\]
The pair $(x^*,y^*)$ is a \emph{Nash equilibrium} if it is a $0$-approximate Nash equilibrium, in which case $\langle x^*,Ay^*\rangle$ is called the value of the game.
A strategy is pure if it is a standard basis vector, and a Nash equilibrium $(x^*,y^*)$ is pure if both $x^*$ and $y^*$ are pure.

It is well known that an approximate Nash equilibrium is obtained as a consequence of no-regret learning dynamics~\citep{freund99adaptive}.
By~\cref{eq:comparator}, the pair of the average strategies $(\bar x_H,\bar y_H)$ is a $\bigl((\Reg_H^x+\Reg_H^y)/H\bigr)$-approximate Nash equilibrium.
Conversely, if $(\bar x_H,\bar y_H)$ is an $\varepsilon$-approximate Nash equilibrium, then~\cref{eq:comparator} gives $\Reg_H^x+\Reg_H^y\le2\varepsilon H$.

\paragraph{Learning dynamics.}
We refer to a pair of learning algorithms, one for each player, as a learning dynamic.
Our lower bounds hold for the following class of learning dynamics, which we call \emph{deterministic} learning dynamics.
In a deterministic learning dynamic, both algorithms for the $x$- and $y$-players can select their strategies at each round $t$ as functions of $m$, $n$, $T$, and all information in the past history $(x_s,y_s,g_s,\ell_s)_{s=1}^{t-1}$.
Thus each player may also use the strategies and the feedback of the opponent, but neither algorithm has any other information about $A$ or uses internal randomization.
The learning dynamics based on optimistic Hedge, in which each player uses only its own feedback, are deterministic learning dynamics.

\section{Main Results}\label{sec:main}
This section presents our regret lower bounds.
Our first result is a lower bound that increases linearly in the number of rounds $H$ up to a parameter $k$, where $k$ is any positive integer with $k\le T$ and $k2^k+1\le\min\{m,n\}$.

\begin{theorem}\label{thm:basic}
Suppose that $k,T,m,n$ are positive integers satisfying $k\le T$ and
$m,n\ge k2^k+1$.
Then, for every deterministic learning dynamic, there exists a fixed payoff matrix
$A\in[-1,1]^{m\times n}$ with a pure Nash equilibrium of value zero such that, for every $H\in[T]$,
\begin{equation}\label{eq:basic-bound}
\Reg_H^x+\Reg_H^y\ge\frac13\min\{H,k\},
\qquad
\max\{\Reg_H^x,\Reg_H^y\}\ge\frac16\min\{H,k\}.
\end{equation}
\end{theorem}

The proof is provided in~\cref{sec:construction,sec:proof-basic}.
In~\cref{thm:basic}, the matrix $A$ depends on the learning dynamic, $k$, and $T$, but the same matrix satisfies~\cref{eq:basic-bound} for every $H\in[T]$.
This matrix has a pure Nash equilibrium, and this implies that the lower bound does not come from the difficulty of finding a mixed equilibrium.

\begin{remark}
The second inequality in~\cref{eq:basic-bound} is a lower bound on the maximum of the two individual regrets, and it is not possible to prove a similar regret lower bound on the \emph{minimum} of the two individual regrets.
In fact, suppose that the $y$-player selects the uniform strategy $\one/n$ in every round, and that the $x$-player selects $\one/m$ in the first round and, in all later rounds, an action that maximizes the first feedback vector $g_1=A\one/n$.
Then $\Reg_H^x=\max_{i\in[m]}g_1(i)-\langle\one/m,g_1\rangle\le2$ for every $A$ and every $H$, which is independent of $k$ and thus $m,n$.
\end{remark}

Choosing $k$ proportional to $\log\min\{m,n\}$ in~\cref{thm:basic} gives the following corollary, which is proved in~\cref{sec:pure-comparisons}.

\begin{corollary}\label{cor:logarithmic}
Suppose that $T\ge1$ and $m,n\ge2$.
Then, for every deterministic learning dynamic, there exists a fixed payoff matrix
$A\in[-1,1]^{m\times n}$ such that, for every $H\in[T]$,
\[
\Reg_H^x+\Reg_H^y
\ge\frac1{12}\min\{H,\log_2\min\{m,n\}\},
\quad
\max\{\Reg_H^x,\Reg_H^y\}
\ge\frac1{24}\min\{H,\log_2\min\{m,n\}\}.
\]
\end{corollary}

When $m=n$, \cref{cor:logarithmic} matches, up to a constant factor, the upper bound $O(\log(mn))$ on the individual regrets of optimistic Hedge with constant learning rates, which holds for every $T$~\citep{rakhlin13optimization}.
Since the individual regrets of every learning dynamic are at most $2T$, the minimax value of the maximum individual regret for deterministic learning dynamics is $\Theta(\min\{T,\log m\})$ for $m=n$.

\Cref{cor:logarithmic} improves the lower bound of~\citet{daskalakis11near} on the rate at which the average strategies converge to a Nash equilibrium.
Combining the lower bound on the social regret in~\cref{cor:logarithmic} with~\cref{eq:comparator} gives the following corollary, which is proved in~\cref{sec:pure-comparisons}.

\begin{corollary}\label{cor:convergence}
Suppose that $T\ge1$ and $m,n\ge2$.
Then, for every deterministic learning dynamic, there exists a fixed payoff matrix
$A\in[-1,1]^{m\times n}$ such that, for every $H\in[T]$, the pair of the average strategies $(\bar x_H,\bar y_H)$ is not an $\varepsilon$-approximate Nash equilibrium of $A$ for any
\[
\varepsilon<\frac1{24H}\min\{H,\log_2\min\{m,n\}\}.
\]
\end{corollary}

For $T\ge\log_2\min\{m,n\}$, the lower bound on $\varepsilon$ in~\cref{cor:convergence} with $H=T$ is $\log_2\min\{m,n\}/(24T)$, whereas the lower bound $\Omega(1/T)$ of~\citet{daskalakis11near} for strongly-uncoupled learning dynamics does not depend on the numbers of actions.
When $m=n$, the lower bound in~\cref{cor:convergence} matches the rate $O(\log(mn)/T)$ of optimistic Hedge with constant learning rates up to a constant factor.

We next consider games in which the two players have different numbers of actions.
The following theorem replaces the condition $n\ge k2^k+1$ in~\cref{thm:basic} by $n\ge16k$, at the cost of a factor $c\le1/4$ in the lower bound.

\begin{theorem}\label{thm:rectangular}
Suppose that $k,T,m,n$ are positive integers satisfying $k\le T$,
$m\ge k2^k+1$, and $n\ge16k$, and define
\begin{equation}\label{eq:rectangular-c}
c=\min\left\{\frac14,
    \sqrt{\frac1{16k}\log\frac{n}{k}}\right\}.
\end{equation}
Then, for every deterministic learning dynamic, there exists
a fixed matrix $A\in[-1,1]^{m\times n}$ with a Nash equilibrium of value zero whose
$x$-player strategy is pure such that, for every $H\in[T]$,
\begin{equation}\label{eq:rectangular-bound}
\Reg_H^x+\Reg_H^y\ge\frac c3\min\{H,k\},
\qquad
\max\{\Reg_H^x,\Reg_H^y\}\ge\frac c6\min\{H,k\}.
\end{equation}
In particular, for every $H$ with $k\le H\le T$,
\begin{equation}\label{eq:rectangular-bound-k}
  \Reg_H^x+\Reg_H^y
  \ge
  \frac1{12}\min\left\{k,
    \sqrt{k\log\frac{n}{k}}\right\},
  \qquad
  \max\{\Reg_H^x,\Reg_H^y\}
  \ge
  \frac1{24}\min\left\{k,
    \sqrt{k\log\frac{n}{k}}\right\}.
\end{equation}
\end{theorem}

The proof is provided in~\cref{sec:proof-rectangular}.
Interchanging the players and replacing $A$ by $-A^\top$ gives the corresponding result with the conditions on $m$ and $n$ exchanged, in which case the $y$-player's equilibrium strategy is pure.
Choosing $k$ proportional to $\log m$ in~\cref{thm:rectangular} gives the following corollary, which is proved in~\cref{sec:proof-rectangular-thm}.

\begin{corollary}\label{cor:rectangular}
For every $\alpha\in(0,1)$, there exists a constant $C_\alpha>0$ such that the following holds.
Suppose that $m\ge n$, $\log m\le n^\alpha$, $T\ge\log_2m$, and $n$ is sufficiently large depending only on $\alpha$.
Then, for every deterministic learning dynamic, there exists a fixed payoff matrix
$A\in[-1,1]^{m\times n}$ such that
\[
\Reg_T^x+\Reg_T^y
\ge2C_\alpha\sqrt{\log m\log n},
\qquad
\max\{\Reg_T^x,\Reg_T^y\}
\ge C_\alpha\sqrt{\log m\log n}.
\]
\end{corollary}

\Cref{cor:rectangular} matches, up to a constant factor, the upper bound $O(\sqrt{\log m\log n})$ on the maximum individual regret, which is achieved by optimistic Hedge when each player knows the number of actions of the opponent~\citep[Theorem~9]{tsuchiya25tight}.
Hence, under the assumptions of~\cref{cor:rectangular}, the minimax value of the maximum individual regret for deterministic learning dynamics is given by
\[
\inf_{(\mathsf{Alg^x},\mathsf{Alg^y})}\ \sup_{A\in[-1,1]^{m\times n}}\ \max\{\Reg_T^x,\Reg_T^y\}=\Theta\bigl(\sqrt{\log m\log n}\bigr),
\]
where the infimum is taken over all deterministic learning dynamics $(\mathsf{Alg^x},\mathsf{Alg^y})$, and the constants in $\Theta$ depend only on $\alpha$.

\begin{remark}
The lower bounds in~\cref{cor:rectangular} do not extend to all $m\ge n$.
In fact, when the number of actions of the $y$-player is $n=2$, there is a deterministic learning dynamic whose individual regrets are both at most $4$ for every $A\in[-1,1]^{m\times2}$ and every $H$~(see~\cref{prop:two-columns} in~\cref{app:two-columns}), whereas $\sqrt{\log m\log n}$ grows with $m$.
\end{remark}

\begin{algorithm}[t]
\caption{Template for constructing the payoff matrix}\label{alg:construction}
\KwIn{a deterministic learning dynamic $(\mathsf{Alg^x}, \mathsf{Alg^y})$, the numbers of actions $m,n$, the horizon $T$, and a number of rounds $k\le T$}
Set $u_0=\one$ and $A_0=0\in\R^{m\times n}$\;
\For{$t=1,\ldots,k$}{
  Let the algorithms $(\mathsf{Alg^x}, \mathsf{Alg^y})$ select $x_t\in\Delta_m$ and $y_t\in\Delta_n$ from $(x_s,y_s,g_s,\ell_s)_{s=1}^{t-1}$\nllabel{line:select}\;
  Choose $u_t\in[-1,1]^m$ with $\langle x_s,u_t\rangle=0$ for all $s\in[t]$\nllabel{line:u}\;
  Choose $v_t\in[-1,1]^n$ with $\langle y_s,v_t\rangle=0$ for all $s\in[t]$\nllabel{line:v}\;
  Set $A_t=A_{t-1}+\frac13(u_t-u_{t-1})v_t^\top$ as in \cref{eq:At_intro}\nllabel{line:update}\;
  Give the feedback $g_t=A_ty_t$ and $\ell_t=A_t^\top x_t$ to $(\mathsf{Alg^x}, \mathsf{Alg^y})$ \nllabel{line:feedback}\;
}
\KwOut{$A=A_k$}
\end{algorithm}

\section{Construction of the Payoff Matrix}\label{sec:construction}
This section gives the construction of the payoff matrix, which is common to the proofs of~\cref{thm:basic,thm:rectangular}.
We construct the matrix from vectors $u_t$ and $v_t$ for \cref{eq:At_intro} that are chosen in each round~(\cref{alg:construction}) and show that the lower bounds hold if these vectors satisfy three conditions~(\cref{lem:matrix}).
The vectors $u_t, v_t$ are then chosen in~\cref{sec:proof-basic} for~\cref{thm:basic} and in~\cref{sec:proof-rectangular} for~\cref{thm:rectangular}.

\paragraph{Required properties of the payoff matrix.}
By~\cref{eq:comparator}, every action $i_*\in[m]$ and every distribution $y^*\in\Delta_n$ satisfy
\begin{equation}\label{eq:star}
\Reg_H^x+\Reg_H^y
\ge\sum_{t=1}^H\bigl(\langle e_{i_*},Ay_t\rangle-\langle x_t,Ay^*\rangle\bigr).
\end{equation}
We will construct a matrix $A$ and choose $i_*$, $y^*$, and a constant $c>0$ satisfying
\begin{equation}\label{eq:goal}
e_{i_*}^\top A=0,
\qquad
Ay^*\le0,
\qquad
\langle x_t,Ay^*\rangle=-\frac c3\quad\text{for all }t\le k,
\end{equation}
where the inequality holds coordinatewise.
Then, under these conditions, continuing from~\cref{eq:star}, we have
\begin{equation}\label{eq:star_plus}
\Reg_H^x+\Reg_H^y
\ge - \sum_{t=1}^H \langle x_t,Ay^*\rangle
= - \sum_{t=1}^{\min\{H,k\}} \langle x_t,Ay^*\rangle - \sum_{t=\min\{H,k\}+1}^H \langle x_t,Ay^*\rangle
\ge \frac{c}{3} \min\{H,k\},
\end{equation}
which is the lower bound on the social regret in~\cref{thm:basic}, where $c=1$, and in~\cref{thm:rectangular}.
The first two properties in~\cref{eq:goal} also imply that the pair $(e_{i_*},y^*)$ is a Nash equilibrium of value zero.
The difficulty is that the strategies $x_t$ in~\cref{eq:goal} are selected by the algorithms and depend on $A$ through the feedback.

\begin{table}[t]
\centering
\caption{Summary of the conditions on the vectors $u_t$ and $v_t$ and of the property of the matrix $A=A_k$ that follows from each condition, where $i_*\in[m]$, $y^*\in\Delta_n$, and $c>0$.
Condition (a) is required in~\cref{line:u,line:v} of~\cref{alg:construction}, and conditions (b)--(d) are the assumptions of~\cref{lem:matrix}.
Under condition (a), the properties in rows (c) and (d) give~\cref{eq:goal}.}
\label{tab:conditions}
\small
\renewcommand{\arraystretch}{1.15}
\begin{tabular}{lll}
\toprule
& Condition on vectors $u_t$ and $v_t$ & Property of payoff matrix $A$\\
\midrule
(a) & $\langle x_s,u_t\rangle=0$ and $\langle y_s,v_t\rangle=0$ for all $1\le s\le t\le k$ & $Ay_s=g_s$ and $A^\top x_s=\ell_s$ for all $s\in[k]$\\
(b) & $\sum_{t=1}^k|u_t(i)-u_{t-1}(i)|\le3$ for all $i\in[m]$ & $A\in[-1,1]^{m\times n}$\\
(c) & $u_t(i_*)=1$ for all $t\in\{0,\ldots,k\}$ & $e_{i_*}^\top A=0$\\
(d) & $\langle y^*,v_t\rangle=c$ for all $t\in[k]$ & $Ay^*=\frac c3(u_k-\one)$\\
\bottomrule
\end{tabular}
\end{table}

\paragraph{Construction.}
Since the strategies depend on $A$, we construct $A$ adaptively while running the two algorithms for $k$ rounds, as described in~\cref{alg:construction}.
At round $t$, the algorithms of the two players select $(x_t,y_t)$ from the strategies and the feedback of the previous rounds~(\cref{line:select}).
The construction then chooses vectors $u_t\in[-1,1]^m$ and $v_t\in[-1,1]^n$ that are orthogonal to the strategies selected so far~(\cref{line:u,line:v}).
Then, it adds the term $\frac13(u_t-u_{t-1})v_t^\top$ to the matrix~(\cref{line:update}) and gives the algorithms the feedback based on the resulting matrix $A_t$~(\cref{line:feedback}).
The payoff matrix is the final matrix $A=A_k$.

The feedback of round $t$ is thus computed from $A_t$, whereas the regrets are defined for the algorithms run on the fixed matrix $A$.
The orthogonality in~\cref{line:u,line:v} is imposed so that the terms added after round $t$ do not change the feedback of round $t$, which implies that the deterministic algorithms select the same strategies when they are run on $A$.
The differences $u_t-u_{t-1}$ in \cref{eq:At_intro} are used to obtain the properties in~\cref{eq:goal}.
If $u_t(i_*)=1$ for all $t$, then the row $i_*$ of every term is zero, and if $\langle y^*,v_t\rangle=c$ for all $t$, then $Ay^*=\frac c3\sum_{t=1}^k(u_t-u_{t-1})=\frac c3(u_k-\one)$.
Since $u_k\le\one$ and the orthogonality in~\cref{line:u} gives $\langle x_t,u_k\rangle=0$ for all $t\le k$, this identity implies $Ay^*\le0$ and $\langle x_t,Ay^*\rangle=-\frac c3$ for all $t\le k$.
The following lemma shows that the lower bounds hold under these two conditions and the condition that $\sum_{t=1}^k|u_t(i)-u_{t-1}(i)|\le3$ for every $i\in[m]$, which ensures that the entries of $A$ are in $[-1,1]$.
\Cref{tab:conditions} summarizes the conditions on the vectors $u_t$ and $v_t$ and the property of $A$ that follows from each of them.
\Cref{alg:construction} does not specify the vectors $u_t$ and $v_t$ beyond the orthogonality, and the lemma applies to every choice that satisfies its assumptions.

\begin{lemma}\label{lem:matrix}
Suppose that $k\le T$ and that the first $k$ rounds are generated by~\cref{alg:construction} for a deterministic learning dynamic.
We also assume that $\sum_{t=1}^k|u_t(i)-u_{t-1}(i)|\le3$ for every $i\in[m]$.
Then the following hold:
\begin{enumerate}[label=(\roman*),ref=(\roman*),leftmargin=*]
\item\label{item:bounded}
It holds that $A_ky_s=g_s$ and $A_k^\top x_s=\ell_s$ for every $s\in[k]$.
Moreover, $A=A_k$ is in the set $[-1,1]^{m\times n}$, and the algorithms run on the fixed game $A$ select $x_1,\ldots,x_k$ and $y_1,\ldots,y_k$ and receive $g_1,\ldots,g_k$ and $\ell_1,\ldots,\ell_k$ in the first $k$ rounds.
\item\label{item:regret}
Further suppose that there exist $i_*\in[m]$ with $u_t(i_*)=1$ for all $t\in\{0,\ldots,k\}$, and $y^*\in\Delta_n$ and $c>0$ with $\langle y^*,v_t\rangle=c$ for all $t\in[k]$.
Then the regrets of the algorithms run on the fixed game $A$ satisfy, for every $H\in[T]$,
\[
\Reg_H^x+\Reg_H^y\ge\frac{c}{3}\min\{H,k\},
\qquad
\max\{\Reg_H^x,\Reg_H^y\}\ge\frac{c}{6}\min\{H,k\},
\]
and $(e_{i_*},y^*)$ is a Nash equilibrium of $A$ with value zero.
\end{enumerate}
\end{lemma}

\begin{proof}
We prove part~\ref{item:bounded} in three steps and then part~\ref{item:regret}.

We first show that $A_ky_s=g_s$ and $A_k^\top x_s=\ell_s$ for every $s\in[k]$.
Fix $s\in[k]$.
By~\cref{line:update,line:feedback} of~\cref{alg:construction},
\[
g_s=A_sy_s,
\qquad
\ell_s=A_s^\top x_s,
\qquad
A_k-A_s=\frac13\sum_{r=s+1}^k(u_r-u_{r-1})v_r^\top.
\]
For every $r\in\{s+1,\ldots,k\}$, we have
\[
(u_r-u_{r-1})v_r^\top y_s
=0,
\qquad
v_r(u_r-u_{r-1})^\top x_s
=\bigl(\langle u_r,x_s\rangle-\langle u_{r-1},x_s\rangle\bigr)v_r
=0,
\]
where the first identity follows from $\langle y_s,v_r\rangle=0$~(\cref{line:v}) and the second from $\langle x_s,u_r\rangle=\langle x_s,u_{r-1}\rangle=0$~(\cref{line:u} at rounds $t=r$ and $t=r-1$, which applies since $s\le r-1$).
Hence $(A_k-A_s)y_s=0$ and $(A_k-A_s)^\top x_s=0$, which gives $A_ky_s=A_sy_s=g_s$ and $A_k^\top x_s=A_s^\top x_s=\ell_s$.

We next show that $A_k\in[-1,1]^{m\times n}$.
For all $i\in[m]$ and $j\in[n]$,
\[
|A_k(i,j)|
\le\frac13\sum_{r=1}^k|u_r(i)-u_{r-1}(i)|\,|v_r(j)|
\le\frac13\sum_{r=1}^k|u_r(i)-u_{r-1}(i)|
\le1,
\]
where the second inequality follows from $v_r\in[-1,1]^n$ and the last from the assumption that $\sum_{r=1}^k|u_r(i)-u_{r-1}(i)|\le3$ for every $i\in[m]$.

We finally show by induction on $s\in[k]$ that the algorithms run on the fixed game $A=A_k$ select $(x_r,y_r)$ and receive $(g_r,\ell_r)$ in every round $r\le s$.
For $s=1$, the algorithms select their strategies at round $1$ as functions of $m$, $n$, and $T$, and hence they select $(x_1,y_1)$ at round $1$ on $A$.
The feedback of round $1$ on $A$ is $Ay_1=g_1$ and $A^\top x_1=\ell_1$ by the first step, and the claim holds for $s=1$.
Let $s\in\{2,\ldots,k\}$, and assume as the induction hypothesis that the claim holds for $s-1$.
By the induction hypothesis, the strategies and the feedback of rounds $1,\ldots,s-1$ on $A$ are those of~\cref{alg:construction}.
Since the algorithms select their strategies at round $s$ as functions of $m$, $n$, $T$, and the strategies and the feedback of rounds $1,\ldots,s-1$, they select $(x_s,y_s)$ at round $s$ on $A$.
The feedback of round $s$ on $A$ is $Ay_s=g_s$ and $A^\top x_s=\ell_s$ by the first step, and the claim holds for $s$.
This proves part~\ref{item:bounded}.

For part~\ref{item:regret}, the definition of $A$ and the assumptions that $u_r(i_*)=1$ for $r\in\{0,\ldots,k\}$ and $\langle y^*,v_r\rangle=c$ for $r\in[k]$ give
\begin{align}
e_{i_*}^\top A
&=\frac13\sum_{r=1}^k\bigl(u_r(i_*)-u_{r-1}(i_*)\bigr)v_r^\top
=0,\label{eq:row}\\
Ay^*
&=\frac13\sum_{r=1}^k\langle v_r,y^*\rangle(u_r-u_{r-1})
=\frac c3\sum_{r=1}^k(u_r-u_{r-1})
=\frac c3(u_k-\one).\label{eq:column}
\end{align}
Since $u_k\le\one$ coordinatewise, \cref{eq:column} gives $Ay^*\le0$ coordinatewise, and hence
\[
\langle x,Ay^*\rangle\le0=\langle e_{i_*},Ay^*\rangle=\langle e_{i_*},Ay\rangle
\qquad\text{for all }x\in\Delta_m,\ y\in\Delta_n,
\]
where the two equalities follow from~\cref{eq:row}.
This shows that $(e_{i_*},y^*)$ is a Nash equilibrium of $A$ with value zero.

It remains to prove the lower bounds on the regrets.
For every round $t\in[T]$ of the algorithms run on $A$, from~\cref{eq:row,eq:column}, each term in the right-hand side of~\cref{eq:star} can be written as
\begin{equation}\label{eq:comparison-gap}
\langle e_{i_*},Ay_t\rangle-\langle x_t,Ay^*\rangle
=0 - \Bigl\langle x_t, \frac c3(u_k-\one) \Bigr\rangle
=\frac c3\bigl(1-\langle x_t,u_k\rangle\bigr)
\geq 0,
\end{equation}
where we used $\langle x_t,\one\rangle=1$ and the last inequality follows from $x_t\in\Delta_m$ and $u_k\le\one$.
The quantity in~\cref{eq:comparison-gap} is equal to $c/3$ for every $t\le k$, since $x_t$ is then the strategy selected in~\cref{alg:construction} by part~\ref{item:bounded}, and~\cref{line:u} of~\cref{alg:construction} gives $\langle x_t,u_k\rangle=0$.
Therefore, using these two properties, for every $H\in[T]$ we have
\[
\Reg_H^x+\Reg_H^y
\ge\sum_{t=1}^H\bigl(\langle e_{i_*},Ay_t\rangle-\langle x_t,Ay^*\rangle\bigr)
=\frac c3\sum_{t=1}^H\bigl(1-\langle x_t,u_k\rangle\bigr)
\ge\frac c3\min\{H,k\},
\]
where the first inequality follows from~\cref{eq:star} and the equality from~\cref{eq:comparison-gap}.
The lower bound on the maximum individual regret follows from $\max\{\Reg_H^x,\Reg_H^y\}\ge(\Reg_H^x+\Reg_H^y)/2$, which completes the proof.
\end{proof}

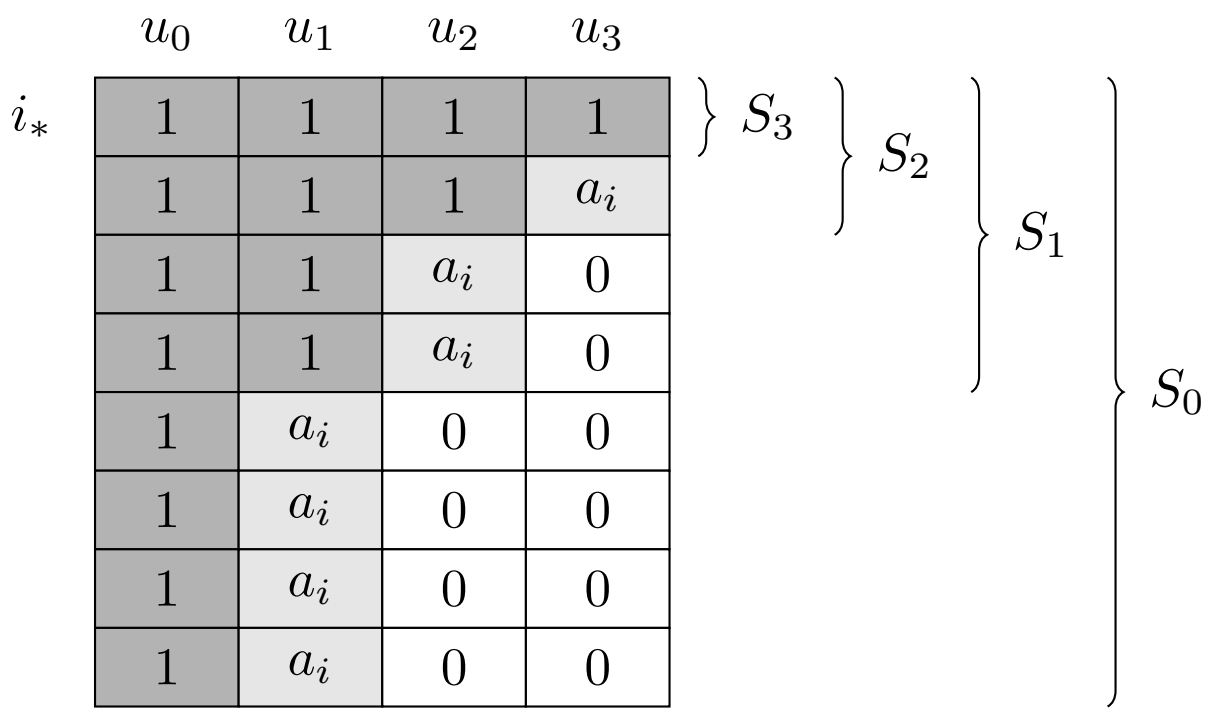
\begin{figure}[t]
\centering
\begin{tikzpicture}[x=0.95cm,y=0.52cm,font=\small]
\tikzset{one/.style={fill=black!30},exit/.style={fill=black!10},zero/.style={fill=white}}
\foreach \t in {0,1,2,3}{\node at (\t+0.5,0.55) {$u_{\t}$};}
\foreach \t in {0,1,2,3}{\draw[one] (\t,0) rectangle ++(1,-1);\node at (\t+0.5,-0.5) {$1$};}
\foreach \t in {0,1,2}{\draw[one] (\t,-1) rectangle ++(1,-1);\node at (\t+0.5,-1.5) {$1$};}
\draw[exit] (3,-1) rectangle ++(1,-1);\node at (3.5,-1.5) {$a_i$};
\foreach \r in {2,3}{\foreach \t in {0,1}{\draw[one] (\t,-\r) rectangle ++(1,-1);\node at (\t+0.5,-\r-0.5) {$1$};}\draw[exit] (2,-\r) rectangle ++(1,-1);\node at (2.5,-\r-0.5) {$a_i$};\draw[zero] (3,-\r) rectangle ++(1,-1);\node at (3.5,-\r-0.5) {$0$};}
\foreach \r in {4,5,6,7}{\draw[one] (0,-\r) rectangle ++(1,-1);\node at (0.5,-\r-0.5) {$1$};\draw[exit] (1,-\r) rectangle ++(1,-1);\node at (1.5,-\r-0.5) {$a_i$};\foreach \t in {2,3}{\draw[zero] (\t,-\r) rectangle ++(1,-1);\node at (\t+0.5,-\r-0.5) {$0$};}}
\node[anchor=east] at (-0.15,-0.5) {$i_*$};
\draw[decorate,decoration={brace,amplitude=3pt}] (4.2,0) -- (4.2,-1) node[midway,right=4pt] {$S_3$};
\draw[decorate,decoration={brace,amplitude=3pt}] (5.15,0) -- (5.15,-2) node[midway,right=4pt] {$S_2$};
\draw[decorate,decoration={brace,amplitude=3pt}] (6.1,0) -- (6.1,-4) node[midway,right=4pt] {$S_1$};
\draw[decorate,decoration={brace,amplitude=3pt}] (7.05,0) -- (7.05,-8) node[midway,right=4pt] {$S_0$};
\end{tikzpicture}
\caption{Illustration of the vectors $u_0,\ldots,u_3$ constructed in~\cref{sec:proof-basic} for $k=3$.
Each row shows the sequence $u_0(i),\ldots,u_3(i)$ of one coordinate $i$, where $a_i\in[-1,1)$.
The set $S_t$ consists of the coordinates that are equal to $1$ in $u_0,\ldots,u_t$, and the coordinate $i_*$ belongs to $S_3$.}
\label{fig:construction}
\end{figure}

By~\cref{lem:matrix}, it remains to choose the vectors $u_t$ and $v_t$ in~\cref{line:u,line:v} of~\cref{alg:construction} so that the assumptions of the lemma hold for some $i_*$, $y^*$, and $c$.
\Cref{sec:proof-basic} gives such vectors with $c=1$ when $m,n\ge k2^k+1$, and \cref{sec:proof-rectangular} gives vectors $v_t$ with the constant $c$ in~\cref{eq:rectangular-c} when $n\ge16k$.

\section{Proof of \texorpdfstring{\cref{thm:basic}}{Theorem 1}}\label{sec:proof-basic}
This section constructs the vectors $u_t$ and $v_t$ in~\cref{line:u,line:v} of~\cref{alg:construction} and proves~\cref{thm:basic}.
We first provide the overview of this section.
The vectors $u_t$ must satisfy $\langle x_s,u_t\rangle=0$ for all $1\le s\le t\le k$, as required in~\cref{line:u}, and \cref{lem:matrix} assumes that $\sum_{t=1}^k|u_t(i)-u_{t-1}(i)|\le3$ for every $i\in[m]$ and that $u_t(i_*)=1$ for all $t\in\{0,\ldots,k\}$ for some $i_*\in[m]$.

To satisfy the condition $\sum_{t=1}^k|u_t(i)-u_{t-1}(i)|\le3$, we let each coordinate change its value at most twice.
Not all coordinates of $u_t$ can be equal to $1$, since $\langle x_t,u_t\rangle=0$ and $x_t\in\Delta_m$.
Coordinates must therefore change from $1$ to other values, and once a coordinate takes a value $a\in[-1,1)$, we set it to zero in all later rounds.
Every sequence $u_0(i),\ldots,u_k(i)$ is then either constantly $1$ or of the form $1,\ldots,1,a,0,\ldots,0$, as illustrated in~\cref{fig:construction}, and in the latter case, the cumulative variation of $u_t(i)$ is upper bounded as
\[
\sum_{t=1}^k|u_t(i)-u_{t-1}(i)|\le|a-1|+|0-a|\le3.
\]

It remains to ensure that some coordinate is equal to $1$ in all rounds.
We define the sets
\[
S_0=[m],
\qquad
S_t=\{i\in S_{t-1}:u_t(i)=1\}\quad\text{for }t\in[k],
\]
which consist of the coordinates that are equal to $1$ in $u_0,\ldots,u_t$.
At round $t$, the coordinates of $u_t$ outside $S_{t-1}$ are zero, and the values $u_t(i)$ for $i\in S_{t-1}$ must satisfy the $t$ linear constraints $\langle x_s,u_t\rangle=0$ for $s\in[t]$.
In~\cref{sec:cube}, we will show that the values $u_t(i)$ for $i\in S_{t-1}$ can be chosen in $[-1,1]$ so that these $t$ constraints hold and at least $(|S_{t-1}|-t)/2$ of the values are equal to $1$~(\cref{lem:cube}), which gives $|S_t|\ge(|S_{t-1}|-t)/2$.
By induction, $|S_k|\ge m/2^k-k$, and hence $S_k$ is nonempty if $m\ge k2^k+1$ as we will see in \cref{lem:nested} in~\cref{sec:nested}.
This halving in each round is the reason why the lower bound in~\cref{thm:basic} increases only for about $\log_2\min\{m,n\}$ rounds.
The vectors $v_t$ are obtained by the same construction applied to the strategies of the $y$-player, and \cref{thm:basic,cor:logarithmic,cor:convergence} are proved in~\cref{sec:pure-comparisons}.

\subsection{Vectors in the cube under linear constraints}\label{sec:cube}
The construction of the vectors $u_t$ is based on the following lemma, which gives a vector in $[-1,1]^d$ that is orthogonal to given vectors and has many coordinates equal to $1$.

\begin{lemma}\label{lem:cube}
Let $a_1,\ldots,a_r\in\R^d$.
Then there exists $z\in[-1,1]^d$ such that $\langle a_s,z\rangle=0$ for all $s\in[r]$ and at most $r$ coordinates of $z$ lie strictly between $-1$ and $1$.
Consequently, there exists such a vector with at least $(d-r)/2$ coordinates equal to $1$.
\end{lemma}

\begin{proof}
Let $P=\{z\in[-1,1]^d:\langle a_s,z\rangle=0\text{ for all }s\in[r]\}$.
Since $P$ contains the zero vector and is compact, there exists
\[
z^*\in\argmax_{z\in P}\|z\|_2^2.
\]
We show that $z^*$ satisfies the first claim.
Define
\[
F=\{i\in[d]:-1<z^*(i)<1\},
\]
and suppose, for contradiction, that $|F|>r$.
Then the $|F|$ vectors $(a_1(i),\ldots,a_r(i))\in\R^r$ for $i\in F$ are linearly dependent, since $|F|>r$.
Hence there exists a nonzero vector $h\in\R^d$ such that
\[
h(i)=0\quad\text{for all }i\notin F,
\qquad
\langle a_s,h\rangle=\sum_{i\in F}h(i)\,a_s(i)=0\quad\text{for all }s\in[r].
\]
Since $-1<z^*(i)<1$ for $i\in F$ and $h(i)=0$ for $i\notin F$, there exists $\delta>0$ such that both $z^*+\delta h$ and $z^*-\delta h$ belong to $[-1,1]^d$.
These two vectors belong to $P$ since $\langle a_s,z^*\pm\delta h\rangle=0$ for all $s\in[r]$, and they satisfy
\[
\|z^*+\delta h\|_2^2+\|z^*-\delta h\|_2^2
=2\|z^*\|_2^2+2\delta^2\|h\|_2^2
>2\|z^*\|_2^2.
\]
Hence at least one of the two vectors has a larger norm than $z^*$, which contradicts the definition of $z^*$.
Therefore $|F|\le r$, which proves the first claim.

For the second claim, at least $d-r$ coordinates of $z^*$ are equal to $1$ or $-1$ by the first claim.
If at least half of them are equal to $1$, then $z^*$ satisfies the second claim.
Otherwise, at least half of them are equal to $-1$, and $-z^*$ satisfies the second claim, since $\langle a_s,-z^*\rangle=0$ for all $s\in[r]$ and the coordinates of $-z^*$ strictly between $-1$ and $1$ are those in $F$.
This completes the proof.
\end{proof}

\subsection{Construction of the vectors \texorpdfstring{$u_t$}{u\_t}}\label{sec:nested}
The following lemma gives vectors $u_t$ that satisfy the orthogonality required in~\cref{line:u} of~\cref{alg:construction} and the assumptions of~\cref{lem:matrix}.
They are constructed by applying~\cref{lem:cube} at each round to the coordinates that are equal to $1$ in all previous rounds.

\begin{lemma}\label{lem:nested}
Suppose that $k$ and $m$ are positive integers satisfying $m\ge k2^k+1$, and let $u_0=\one$.
Then, for any sequentially given $x_1,\ldots,x_k\in\Delta_m$, one can choose $u_t\in[-1,1]^m$ for $t\in[k]$, with $u_t$ depending only on $x_1,\ldots,x_t$, and a coordinate $i_*\in[m]$ such that
\begin{align}
\langle x_s,u_t\rangle&=0 &&\text{for all }1\le s\le t\le k,\label{eq:nested-orth}\\
\sum_{t=1}^k|u_t(i)-u_{t-1}(i)|&\le3 &&\text{for all }i\in[m],\label{eq:nested-var}\\
u_t(i_*)&=1 &&\text{for all }t\in\{0,\ldots,k\}.\label{eq:nested-one}
\end{align}
\end{lemma}

\begin{proof}
We construct the vectors $u_t$ together with sets $S_t\subseteq[m]$ recursively for $t=1,\ldots,k$, and then prove the three properties in the statement.

Set $S_0=[m]$.
Let $t\in[k]$, and suppose that $u_{t-1}$ and $S_{t-1}$ have been constructed from $x_1,\ldots,x_{t-1}$.
We set $u_t(i)=0$ for $i\notin S_{t-1}$.
For $i\in S_{t-1}$, we choose the values $u_t(i)\in[-1,1]$ by the second claim of~\cref{lem:cube}, applied with $d=|S_{t-1}|$ and $r=t$ to the vectors $(x_s(i))_{i\in S_{t-1}}$ for $s\in[t]$, which are the restrictions of $x_1,\ldots,x_t$ to the coordinates in $S_{t-1}$.
These values satisfy
\begin{equation}\label{eq:restricted}
\sum_{i\in S_{t-1}}x_s(i)u_t(i)=0
\qquad\text{for all }s\in[t],
\end{equation}
and at least $(|S_{t-1}|-t)/2$ of them are equal to $1$.
We then define
\[
S_t=\{i\in S_{t-1}:u_t(i)=1\}.
\]
Then, the lower bound on the number of values equal to $1$ gives
\begin{equation}\label{eq:halving}
|S_t|\ge\frac{|S_{t-1}|-t}2.
\end{equation}
By construction, $u_t$ and $S_t$ depend only on $x_1,\ldots,x_t$, and $S_0\supseteq S_1\supseteq\cdots\supseteq S_k$.

The orthogonality~\cref{eq:nested-orth} directly follows from the above construction.
For $1\le s\le t\le k$, since $u_t(i)=0$ for $i\notin S_{t-1}$, \cref{eq:restricted} gives
\[
\langle x_s,u_t\rangle=\sum_{i\in S_{t-1}}x_s(i)u_t(i)=0.
\]

We next prove~\cref{eq:nested-one} for some $i_*\in[m]$.
To prove this, we first show by induction on $t$ that
\begin{equation}\label{eq:survivors}
|S_t|\ge\frac{m}{2^t}-t
\qquad\text{for every }t\in\{0,\ldots,k\}.
\end{equation}
For $t=0$, \cref{eq:survivors} holds since $|S_0|=m$.
Let $t\in[k]$, and assume as the induction hypothesis that $|S_{t-1}|\ge m/2^{t-1}-(t-1)$.
Then
\[
|S_t|
\ge\frac{|S_{t-1}|-t}2
\ge\frac12\left(\frac{m}{2^{t-1}}-(t-1)-t\right)
=\frac{m}{2^t}-t+\frac12
\ge\frac{m}{2^t}-t,
\]
where the first inequality follows from~\cref{eq:halving} and the second from the induction hypothesis.
Hence \cref{eq:survivors} holds, and the assumption $m\ge k2^k+1$ gives
\[
|S_k|\ge\frac{m}{2^k}-k\ge2^{-k}>0.
\]
Fix $i_*\in S_k$.
Since $S_k\subseteq S_t$, we have $i_*\in S_t$ for every $t\in[k]$, and the definition of $S_t$ gives $u_t(i_*)=1$ for every $t\in[k]$.
Since $u_0=\one$, the coordinate $i_*$ satisfies $u_t(i_*)=1$ for all $t\in\{0,\ldots,k\}$.

It remains to prove~\cref{eq:nested-var}.
Fix $i\in[m]$.
If $i\in S_k$, then $u_t(i)=1$ for all $t\in\{0,\ldots,k\}$ by the same argument as for $i_*$, and the sum in~\cref{eq:nested-var} is zero.
If $i\notin S_k$, define $p=\min\{t\in[k]:i\notin S_t\}$.
Then $i\in S_t$ for $t<p$ and $i\notin S_t$ for $t\ge p$, and hence
\[
u_t(i)=1\quad\text{for }t<p,
\qquad
u_p(i)\in[-1,1),
\qquad
u_t(i)=0\quad\text{for }t>p,
\]
where the first relation follows from $u_0=\one$ and the definition of $S_t$, the second from $i\in S_{p-1}\setminus S_p$ and the definition of $S_p$, and the third from $i\notin S_{t-1}$ and the definition of $u_t$ outside $S_{t-1}$.
The sequence $u_0(i),\ldots,u_k(i)$ is therefore of the form $1,\ldots,1,u_p(i),0,\ldots,0$, and
\[
\sum_{t=1}^k|u_t(i)-u_{t-1}(i)|
\le|u_p(i)-1|+|0-u_p(i)|
\le3,
\]
which completes the proof.
\end{proof}

\subsection{Proofs of \texorpdfstring{\cref{thm:basic,cor:logarithmic,cor:convergence}}{Theorem 1 and Corollaries 1 and 2}}\label{sec:pure-comparisons}
To apply~\cref{lem:matrix}, it remains to choose the vectors $v_t$ in~\cref{line:v} of~\cref{alg:construction} and a distribution $y^*$ with $\langle y^*,v_t\rangle=c$ for all $t\in[k]$.
For~\cref{thm:basic}, we determine both by applying~\cref{lem:nested} to the strategies of the $y$-player and taking $y^*=e_{j_*}$ for a coordinate $j_*$ with $v_t(j_*)=1$ for all $t$, which gives $c=1$.

\begin{proof}[Proof of~\cref{thm:basic}]
Fix any deterministic learning dynamic, and run~\cref{alg:construction} for $k$ rounds.
We choose $u_t$ in~\cref{line:u} by applying~\cref{lem:nested} to $x_1,\ldots,x_t$, and $v_t$ in~\cref{line:v} by applying~\cref{lem:nested} to $y_1,\ldots,y_t$ with $n$ in place of $m$.
This is possible since $m,n\ge k2^k+1$, and the vectors $u_t$ and $v_t$ satisfy the requirements of~\cref{line:u,line:v} by~\cref{eq:nested-orth} in~\cref{lem:nested}.
Both choices use only the strategies selected up to round $t$, and hence the feedback of round $t$ in~\cref{line:feedback} is determined before the algorithms select their strategies at round $t+1$.

We verify the assumptions of~\cref{lem:matrix}.
By~\cref{eq:nested-var} in~\cref{lem:nested}, we have $\sum_{t=1}^k|u_t(i)-u_{t-1}(i)|\le3$ for every $i\in[m]$.
By~\cref{eq:nested-one} in~\cref{lem:nested} in both applications, there exist $i_*\in[m]$ and $j_*\in[n]$ such that $u_t(i_*)=1$ and $v_t(j_*)=1$ for every $t\in\{0,\ldots,k\}$, where $v_0=\one$.
With $y^*=e_{j_*}$ and $c=1$, we have $\langle y^*,v_t\rangle=v_t(j_*)=c$ for every $t\in[k]$.
Hence all assumptions of~\cref{lem:matrix} hold.

Therefore, by~\cref{lem:matrix}, the matrix $A=A_k$ belongs to $[-1,1]^{m\times n}$, the pair $(e_{i_*},e_{j_*})$ is a pure Nash equilibrium of $A$ with value zero, and the regrets of the algorithms run on the fixed game $A$ satisfy the desired regret lower bound~\cref{eq:basic-bound} for every $H\in[T]$, which completes the proof.
\end{proof}

\begin{proof}[Proof of~\cref{cor:logarithmic}]
Write $d=\min\{m,n\}$.
Since $\max\{\Reg_H^x,\Reg_H^y\}\ge(\Reg_H^x+\Reg_H^y)/2$, it suffices to prove the first inequality.
We first suppose that $d\ge4$ and apply~\cref{thm:basic} with
\[
k=\min\bigl\{T,\lfloor\tfrac12\log_2d\rfloor\bigr\},
\]
which satisfies $1\le k\le T$.
Since $2^k\ge k+1$ and $2k\le\log_2d$, we have $k2^k+1\le(k+1)2^k\le4^k\le d$, and hence $m,n\ge k2^k+1$.
Moreover, $\lfloor z\rfloor\ge z/2$ for $z\ge1$ gives $\lfloor\tfrac12\log_2d\rfloor\ge\tfrac14\log_2d$, and hence, whether $k=T$ or $k=\lfloor\tfrac12\log_2d\rfloor$,
\begin{equation}\label{eq:k-log}
\min\{H,k\}\ge\tfrac14\min\{H,\log_2d\}
\qquad\text{for every }H\in[T].
\end{equation}
By the first inequality of~\cref{eq:basic-bound} in~\cref{thm:basic} and~\cref{eq:k-log}, we have
\[
\Reg_H^x+\Reg_H^y
\ge\frac13\min\{H,k\}
\ge\frac1{12}\min\{H,\log_2d\}
\qquad\text{for every }H\in[T],
\]
which proves the claim for $d\ge4$.

It remains to consider the case of $d\in\{2,3\}$.
In this case, we construct a matrix for which $\Reg_H^x+\Reg_H^y\ge1$ for every $H\in[T]$, which is larger than the stated bound since $\frac1{12}\log_2 3<1$.
Since the first strategy $x_1$ does not depend on $A$, we can choose $i\in[m]$ with $x_1(i)\le1/m$ and define $A=(2e_i-\one)\one^\top\in[-1,1]^{m\times n}$.
Then $Ay=2e_i-\one$ for every $y\in\Delta_n$, and hence \cref{eq:star} with $i_*=i$ gives, for every $H\in[T]$,
$
\Reg_H^x+\Reg_H^y
\ge\sum_{t=1}^H (1-\langle x_t,2e_i-\one\rangle)
=\sum_{t=1}^H 2 (1-x_t(i))
\ge2(1-1/m)
\ge1,
$
which completes the proof.
\end{proof}

\begin{proof}[Proof of~\cref{cor:convergence}]
Let $A$ be the matrix of~\cref{cor:logarithmic}, and fix $H\in[T]$.
By~\cref{eq:comparator} and the first inequality in~\cref{cor:logarithmic}, we have
\[
\max_{x\in\Delta_m}\langle x,A\bar y_H\rangle-\min_{y\in\Delta_n}\langle\bar x_H,Ay\rangle
=\frac{\Reg_H^x+\Reg_H^y}{H}
\ge\frac1{12H}\min\{H,\log_2\min\{m,n\}\}.
\]
If $(\bar x_H,\bar y_H)$ is an $\varepsilon$-approximate Nash equilibrium of $A$, then the left-hand side is at most $2\varepsilon$, and hence $\varepsilon\ge\frac1{24H}\min\{H,\log_2\min\{m,n\}\}$, which completes the proof.
\end{proof}

\section{Proof of \texorpdfstring{\cref{thm:rectangular}}{Theorem 2}}\label{sec:proof-rectangular}
This section proves~\cref{thm:rectangular}.
We first provide the overview of the proof in this section.
The vectors $u_t$ are constructed as in~\cref{sec:proof-basic}.
For the vectors $v_t$, a coordinate $j_*$ with $v_t(j_*)=1$ for all $t\in[k]$ need not exist when $n<k2^k+1$.
Still, noticing that the assumption $\langle y^*,v_t\rangle=c$ in~\cref{lem:matrix} does not require $y^*$ to be a pure strategy,
we can construct the vectors $v_t$ together with a distribution $y^*\in\Delta_n$ that satisfies this assumption for a constant $c\in(0,1/4]$.
Then, by~\cref{lem:matrix}, we can obtain the social regret lower bound of $\frac c3\min\{H,k\}$.

The distribution $y^*$ must satisfy $\langle y^*,v_t\rangle=c$ for all $t\in[k]$, whereas each $v_t$ must be determined at round $t$ from $y_1,\ldots,y_t$, since the feedback $\ell_t=A_t^\top x_t$ depends on $v_t$.
We therefore construct $y^*$ recursively.
At each round $t$, together with $v_t$, we construct a distribution $y^*_t\in\Delta_n$ that satisfies $\langle y^*_t,v_s\rangle=c$ for all $s\le t$, and we set $y^*=y^*_k$.
Starting from the uniform distribution $y^*_0=\one/n$, we choose
\[
y^*_t(j)=y^*_{t-1}(j)\bigl(1+\eta_t\,v_t(j)\bigr)
\qquad\text{for }j\in[n],
\]
where $\eta_t>0$ is a step size.
Since $|v_t(j)|\le1$, this multiplicative update keeps all coordinates positive if $\eta_t<1$.

It remains to choose the vector $v_t$ and the step size $\eta_t$.
The vector $v_t$ is chosen by~\cref{lem:cube} under $2t$ linear constraints, which consist of the orthogonality to $y_1,\ldots,y_t$ required in~\cref{line:v} of~\cref{alg:construction} and $t$ further constraints under which the coordinates of $y^*_t$ sum to $1$ and $\langle y^*_t,v_s\rangle=c$ holds for all $s<t$.
The step size $\eta_t$ is then chosen so that $\langle y^*_t,v_t\rangle=c$.
We will show that $\eta_t\le1/2$ holds for all $t\in[k]$ if $c\le1/4$ and $n\ge8k\exp(4kc^2)$, which are the assumptions of~\cref{lem:rectangular-comparator} in~\cref{sec:mixed-comparator}.
The second condition is equivalent to $4kc^2\le\log(n/(8k))$, which is the reason why the constant $c$ in~\cref{eq:rectangular-c} and the lower bound in~\cref{thm:rectangular} contain the terms $\sqrt{\log(n/k)/k}$ and $\sqrt{k\log(n/k)}$, respectively.
The constant $c$ in~\cref{eq:rectangular-c} satisfies both conditions when $n\ge16k$, and \cref{thm:rectangular,cor:rectangular} are proved in~\cref{sec:proof-rectangular-thm}.

\subsection{Construction of the vectors \texorpdfstring{$v_t$}{v\_t} and the distribution \texorpdfstring{$y^*$}{y*}}\label{sec:mixed-comparator}
The following lemma gives the vectors $v_t$ and the distribution $y^*$ described above, which satisfy the orthogonality required in~\cref{line:v} of~\cref{alg:construction} and the assumption $\langle y^*,v_t\rangle=c$ in~\cref{lem:matrix}.

\begin{lemma}\label{lem:rectangular-comparator}
Suppose that $k,n$ are positive integers and $c$ is a real number satisfying $0<c\le1/4$ and $n\ge8k\exp(4kc^2)$.
Then, for any sequentially given $y_1,\ldots,y_k\in\Delta_n$, one can
choose $v_t\in[-1,1]^n$ depending only on $y_1,\ldots,y_t$ and, after round $k$,
a distribution $y^*\in\Delta_n$ such that
\begin{align}
  \langle y_s,v_t\rangle&=0 &&\text{for all }1\le s\le t\le k,\label{eq:rectangular-orth}\\
  \langle y^*,v_t\rangle&=c &&\text{for all }t\in[k].\label{eq:rectangular-comparator}
\end{align}
\end{lemma}

\begin{proof}
We construct vectors $v_1,\ldots,v_k$ and distributions $y^*_0,\ldots,y^*_k$ recursively, and we show by induction on $t\in\{0,\ldots,k\}$ that
\begin{equation}\label{eq:rectangular-invariant}
  \text{(a)}\ \,y^*_t\in\Delta_n,
  \quad\ \text{(b)}\ \,\langle y^*_t,v_s\rangle=c\text{ for all }s\in[t],
  \quad\ \text{(c)}\ \,\|y^*_t\|_2^2\le\frac{\exp(4tc^2)}n.
\end{equation}
For $t=k$, the conditions~\eqpart{eq:rectangular-invariant}{a} and~\eqpart{eq:rectangular-invariant}{b} state that $y^*=y^*_k$ belongs to $\Delta_n$ and satisfies~\cref{eq:rectangular-comparator}, and~\eqpart{eq:rectangular-invariant}{c} is used only in the induction step.

For $t=0$, we set $y^*_0=\one/n$.
Then~\eqpart{eq:rectangular-invariant}{a} holds, \eqpart{eq:rectangular-invariant}{b} contains no condition, and~\eqpart{eq:rectangular-invariant}{c} holds since $\|y^*_0\|_2^2=1/n$.

Let $t\in[k]$, and assume as the induction hypothesis that $v_1,\ldots,v_{t-1}$ and $y^*_{t-1}$ have been constructed from $y_1,\ldots,y_{t-1}$ and that~\cref{eq:rectangular-invariant} holds with $t-1$ in place of $t$.
We define $y^*_t$ by
\begin{equation}\label{eq:rectangular-update}
  y^*_t(j)=y^*_{t-1}(j)\bigl(1+\eta_t\,v_t(j)\bigr)
  \qquad\text{for }j\in[n],
\end{equation}
where the vector $v_t\in[-1,1]^n$ and the step size $\eta_t>0$ are chosen in what follows.
For every choice of $v_t$ and $\eta_t$, the definition~\eqref{eq:rectangular-update} and the induction hypotheses~\eqpart{eq:rectangular-invariant}{a} and~\eqpart{eq:rectangular-invariant}{b} give
\begin{align}
\sum_{j=1}^n y^*_t(j)
&=1+\eta_t\langle y^*_{t-1},v_t\rangle,\label{eq:rectangular-id-sum}\\
\langle y^*_t,v_s\rangle
&=c+\eta_t\sum_{j=1}^n y^*_{t-1}(j)v_s(j)v_t(j)
\qquad\text{for all }1\le s<t,\label{eq:rectangular-id-prev}\\
\langle y^*_t,v_t\rangle
&=\langle y^*_{t-1},v_t\rangle+\eta_t\sum_{j=1}^n y^*_{t-1}(j)v_t(j)^2,\label{eq:rectangular-id-new}\\
\|y^*_t\|_2^2
&=\|y^*_{t-1}\|_2^2
  +2\eta_t\sum_{j=1}^n y^*_{t-1}(j)^2v_t(j)
  +\eta_t^2\sum_{j=1}^n y^*_{t-1}(j)^2v_t(j)^2.\label{eq:rectangular-id-norm}
\end{align}

We first choose $v_t$ as a vector in $[-1,1]^n$ that satisfies the following $2t$ linear constraints:
\begin{align}
    \langle y_s,v_t\rangle&=0 &&\text{for all }s\in[t],\label{eq:rectangular-con-y}\\
    \langle y^*_{t-1},v_t\rangle&=0,\label{eq:rectangular-con-star}\\
    \sum_{j=1}^n y^*_{t-1}(j)v_s(j)v_t(j)&=0 &&\text{for all }1\le s<t.\label{eq:rectangular-con-prev}
\end{align}
The constraint~\eqref{eq:rectangular-con-y} is the orthogonality required in~\cref{eq:rectangular-orth}.
Under the constraints~\cref{eq:rectangular-con-star,eq:rectangular-con-prev}, the identities~\cref{eq:rectangular-id-sum,eq:rectangular-id-prev} give $\sum_{j=1}^ny^*_t(j)=1$ and $\langle y^*_t,v_s\rangle=c$ for all $s<t$, which matches the requirement of~\eqpart{eq:rectangular-invariant}{a} and~\eqpart{eq:rectangular-invariant}{b} at round $t$.
The constraints~\eqref{eq:rectangular-con-y}--\eqref{eq:rectangular-con-prev} require $v_t$ to be orthogonal to the $2t$ vectors $y_1,\ldots,y_t$, $y^*_{t-1}$, and $(y^*_{t-1}(j)v_s(j))_{j\in[n]}$ for $1\le s<t$.
Hence the first claim of~\cref{lem:cube} with $d=n$ and $r=2t$ gives a vector $v_t\in[-1,1]^n$ that satisfies~\cref{eq:rectangular-con-y}--\cref{eq:rectangular-con-prev} and has at most $2t$ coordinates strictly between $-1$ and $1$.
Since $-v_t$ also has these two properties, we can choose the sign of $v_t$ so that
\begin{equation}\label{eq:rectangular-sign}
  \sum_{j=1}^n y^*_{t-1}(j)^2v_t(j)\le0,
\end{equation}
which makes the term $2\eta_t\sum_{j=1}^ny^*_{t-1}(j)^2v_t(j)$ in~\cref{eq:rectangular-id-norm} nonpositive.

We next choose $\eta_t$ so that $\langle y^*_t,v_t\rangle=c$.
By~\cref{eq:rectangular-id-new} and~\cref{eq:rectangular-con-star}, we have $\langle y^*_t,v_t\rangle=\eta_t\sum_{j=1}^ny^*_{t-1}(j)v_t(j)^2$, and we first bound this sum from below.
By the induction hypothesis~\eqpart{eq:rectangular-invariant}{c}, we have
\[
  \|y^*_{t-1}\|_2^2
  \le\frac{\exp(4(t-1)c^2)}n
  \le\frac{\exp(4kc^2)}n
  \le\frac1{8k}
  \le\frac1{8t},
\]
where the third inequality follows from the assumption $n\ge8k\exp(4kc^2)$, and the second and the fourth from $t\le k$.
Hence
\begin{equation}\label{eq:rectangular-denominator}
    \sum_{j=1}^n y^*_{t-1}(j)v_t(j)^2
    \ge\sum_{j:\,|v_t(j)|=1}y^*_{t-1}(j)
    =1-\sum_{j:\,|v_t(j)|<1}y^*_{t-1}(j)
    \ge1-\sqrt{2t}\,\|y^*_{t-1}\|_2
    \ge\frac12,
\end{equation}
where the first inequality follows from $y^*_{t-1}(j)\ge0$, the equality from $\sum_{j=1}^ny^*_{t-1}(j)=1$, the second inequality from the Cauchy--Schwarz inequality and the fact that at most $2t$ coordinates satisfy $|v_t(j)|<1$, and the last inequality from $\|y^*_{t-1}\|_2^2\le1/(8t)$.
We define
\begin{equation}\label{eq:rectangular-eta}
  \eta_t=\frac{c}{\sum_{j=1}^n y^*_{t-1}(j)v_t(j)^2}.
\end{equation}
By~\cref{eq:rectangular-denominator} and the assumption $c\le1/4$, the step size satisfies $0<\eta_t\le2c\le1/2$.
With these choices of $v_t$ and $\eta_t$, \cref{eq:rectangular-update} defines $y^*_t$, and $v_t$ and $y^*_t$ depend only on $y_1,\ldots,y_t$.

It remains to prove that~\cref{eq:rectangular-invariant} holds at round $t$.
By~\cref{eq:rectangular-con-star}, \cref{eq:rectangular-con-prev}, and~\cref{eq:rectangular-eta}, the identities~\eqref{eq:rectangular-id-sum}--\eqref{eq:rectangular-id-new} give $\sum_{j=1}^ny^*_t(j)=1$ and $\langle y^*_t,v_s\rangle=c$ for all $s\in[t]$.
Every coordinate satisfies $y^*_t(j)\ge y^*_{t-1}(j)(1-\eta_t)\ge0$ by $y^*_{t-1}(j)\ge0$, $|v_t(j)|\le1$, and $\eta_t\le1/2$.
Hence~\eqpart{eq:rectangular-invariant}{a} and~\eqpart{eq:rectangular-invariant}{b} hold.
For~\eqpart{eq:rectangular-invariant}{c}, the identity~\eqref{eq:rectangular-id-norm} gives
\[
  \|y^*_t\|_2^2
  \le\left(1+\eta_t^2\right)\|y^*_{t-1}\|_2^2
  \le\exp(4c^2)\|y^*_{t-1}\|_2^2
  \le\frac{\exp(4tc^2)}n,
\]
where the first inequality follows from~\cref{eq:rectangular-sign} and $|v_t(j)|\le1$, the second from $\eta_t\le2c$ and $1+z\le\exp(z)$, and the third from the induction hypothesis~\eqpart{eq:rectangular-invariant}{c}.
Hence~\cref{eq:rectangular-invariant} holds for every $t\in\{0,\ldots,k\}$.

The vectors $v_t$ satisfy~\cref{eq:rectangular-orth} by~\cref{eq:rectangular-con-y}, and the distribution $y^*=y^*_k$ belongs to $\Delta_n$ and satisfies~\cref{eq:rectangular-comparator} by~\eqpart{eq:rectangular-invariant}{a} and~\eqpart{eq:rectangular-invariant}{b} with $t=k$, which completes the proof.
\end{proof}

\subsection{Proofs of \texorpdfstring{\cref{thm:rectangular,cor:rectangular}}{Theorem 2 and Corollary 3}}\label{sec:proof-rectangular-thm}
We first verify that the constant $c$ in~\cref{eq:rectangular-c} satisfies the assumptions of~\cref{lem:rectangular-comparator} and then apply~\cref{lem:matrix} as in the proof of~\cref{thm:basic}.

\begin{proof}[Proof of~\cref{thm:rectangular}]
\Cref{lem:rectangular-comparator} assumes that $0<c\le1/4$ and that $n\ge8k\exp(4kc^2)$.
The first assumption follows from~\cref{eq:rectangular-c} and $n\ge16k$.
The second assumption is equivalent to $4kc^2\le\log(n/(8k))$, which holds since
\[
  4kc^2
  \le\frac14\log\frac nk
  \le\log\frac nk-\log8
  =\log\frac n{8k},
\]
where the first inequality follows from~\cref{eq:rectangular-c} and the second from $n\ge16k$ and $\frac34\log16=\log8$.

Fix any deterministic learning dynamic, and run~\cref{alg:construction} for $k$ rounds.
We choose $u_t$ in~\cref{line:u} by applying~\cref{lem:nested} to $x_1,\ldots,x_t$, and $v_t$ in~\cref{line:v} by applying~\cref{lem:rectangular-comparator} to $y_1,\ldots,y_t$.
\Cref{lem:nested} applies since $m\ge k2^k+1$, and the vectors $u_t$ and $v_t$ satisfy the requirements of~\cref{line:u,line:v} by~\cref{eq:nested-orth} in~\cref{lem:nested} and~\cref{eq:rectangular-orth} in~\cref{lem:rectangular-comparator}.
Both choices use only the strategies selected up to round $t$, and hence the feedback of round $t$ in~\cref{line:feedback} is determined before the algorithms select their strategies at round $t+1$.

We now verify the assumptions of~\cref{lem:matrix}.
By~\cref{eq:nested-var} in~\cref{lem:nested}, we have $\sum_{t=1}^k|u_t(i)-u_{t-1}(i)|\le3$ for every $i\in[m]$.
By~\cref{eq:nested-one} in~\cref{lem:nested}, there exists $i_*\in[m]$ with $u_t(i_*)=1$ for every $t\in\{0,\ldots,k\}$.
By~\cref{eq:rectangular-comparator} in~\cref{lem:rectangular-comparator}, there exists a distribution $y^*\in\Delta_n$ with $\langle y^*,v_t\rangle=c$ for every $t\in[k]$.
Hence all assumptions of~\cref{lem:matrix} hold.

Therefore, by~\cref{lem:matrix}, the matrix $A=A_k$ belongs to $[-1,1]^{m\times n}$, the pair $(e_{i_*},y^*)$ is a Nash equilibrium of $A$ with value zero, and the regrets of the algorithms run on the fixed game $A$ satisfy~\cref{eq:rectangular-bound} for every $H\in[T]$.
For every $H$ with $k\le H\le T$, \cref{eq:rectangular-bound,eq:rectangular-c} give
\[
\Reg_H^x+\Reg_H^y
\ge\frac{ck}3
=\frac1{12}\min\left\{k,\sqrt{k\log\frac nk}\right\},
\;
\max\{\Reg_H^x,\Reg_H^y\}
\ge\frac{ck}6
=\frac1{24}\min\left\{k,\sqrt{k\log\frac nk}\right\},
\]
which is~\cref{eq:rectangular-bound-k} and completes the proof.
\end{proof}

\begin{proof}[Proof of~\cref{cor:rectangular}]
We take $n\ge16^{1/(1-\alpha)}$ as the condition in the statement that $n$ is sufficiently large, and prove the claim by applying~\cref{thm:rectangular} with $k=\lfloor\tfrac12\log_2m\rfloor$.
Since $m\ge n\ge16$, we have $\tfrac12\log_2m\ge1$, and $z/2\le\lfloor z\rfloor\le z$ for $z\ge1$ gives $\tfrac14\log_2m\le k\le\tfrac12\log_2m$.

We first verify the assumptions $k\le T$, $m\ge k2^k+1$, and $n\ge16k$ of~\cref{thm:rectangular}.
Since $2k\le\log_2m$, we have $k\le\log_2m\le T$ and $m\ge4^k\ge k2^k+1$, where the last inequality follows from $2^k\ge k+1$.
Moreover, $2\log2\ge1$ and the assumption $\log m\le n^\alpha$ give $k\le\tfrac12\log_2m=\frac{\log m}{2\log2}\le\log m\le n^\alpha$, and hence $n/k\ge n^{1-\alpha}\ge16$.
Therefore $n\ge16k$, and \cref{thm:rectangular} applies.

It remains to bound the two terms in the minimum of~\cref{eq:rectangular-bound-k} in~\cref{thm:rectangular} from below.
Since $k\ge\tfrac14\log_2m=\frac{\log m}{4\log2}$, $m\ge n$, and $n/k\ge n^{1-\alpha}$, we have
\[
k\ge\frac{\log m}{4\log2}
\ge\frac{\sqrt{\log m\log n}}{4\log2},
\qquad
\sqrt{k\log\frac{n}{k}}
\ge\sqrt{\frac{\log m}{4\log2}\,(1-\alpha)\log n}
=\sqrt{\frac{1-\alpha}{4\log2}}\,\sqrt{\log m\log n}.
\]
Applying~\cref{eq:rectangular-bound-k} in~\cref{thm:rectangular} with $H=T\ge k$ therefore gives the claim with $C_\alpha=\min\{1,\sqrt{4(1-\alpha)\log2}\}/(96\log2)$, which completes the proof.
\end{proof}

\section{Discussion}\label{sec:discussion}
This section discusses a consequence of~\cref{thm:basic} for the query complexity of approximate Nash equilibria, and an open problem.

\begin{figure}[t]
\centering
\begin{tikzpicture}[x=0.86cm,y=0.76cm,font=\small,>=stealth]
\tikzset{upper/.style={thick,blue!65!black},lower/.style={thick,dashed,red!70!black},new/.style={line width=2pt,red!70!black},lead/.style={black!45,thin}}
\def\ub{blue!65!black}\def\lb{red!70!black}
\draw[->] (0,0) -- (15.9,0) node[right] {$\varepsilon$};
\draw[->] (0,0) -- (0,7);
\node[rotate=90] at (-2.25,3.5) {number of queries};
\foreach \x/\l in {0/$0$,2/$c\,d^{-4}$,11.2/$\frac1{12}$,14.2/$1-\frac1d$,15.3/$1$}{\draw (\x,0.08) -- (\x,-0.08) node[below] {\l};}
\foreach \y/\l in {0.5/$1$,1.75/$\frac1{48}\log_2d$,5.75/$\frac d2-1$,6.4/$d$}{\draw (0.08,\y) -- (-0.08,\y) node[left] {\l};}
\begin{scope}
\clip (7.59,0.5) rectangle (11.2,1.75);
\fill[red!70!black!14] (6.6,3.5) .. controls (7.1,2.2) and (8.6,0.95) .. (11.9,0.5) -- (11.9,3.5) -- cycle;
\end{scope}
\draw[upper] (0,6.4) -- (5.6,6.4) .. controls (6.7,4.45) and (9,2.85) .. (11.2,2.35) .. controls (12.3,2.1) and (13.4,2) .. (14.2,1.98) -- (14.2,0) -- (15.3,0);
\draw[upper] (4.6,6.4) .. controls (5.3,4.95) and (7.3,4.01) .. (9.4,2.99);
\draw[lower] (0,5.75) .. controls (0.25,3.2) and (0.9,0.9) .. (2,0.5);
\draw[lower,densely dotted] (2,0.5) -- (14.2,0.5) -- (14.2,0);
\draw[lower] (0.35,3.5) -- (6.6,3.5) .. controls (7.1,2.2) and (8.6,0.95) .. (11.9,0.5);
\draw[new] (0,1.75) -- (11.2,1.75);
\fill[\lb] (11.2,1.75) circle (2.4pt);
\draw[densely dotted,black!60] (11.2,0) -- (11.2,2.35);
{\hypersetup{citecolor=blue!65!black}
\node[text=\ub,anchor=west,align=left] at (8.7,5.8) {$O(\log d/\varepsilon)$\\[-1pt]{\footnotesize\citep{nemirovski04prox,nesterov05smooth}}};
\draw[lead] (8.85,5.32) -- (7.33,4.51);
\node[text=\ub,anchor=east,align=right] at (4.75,5.76) {$\tilde O(\varepsilon^{-2/3})$\\[-1pt]{\footnotesize\citep{karmarkar26solving}}};
\node[text=\ub,anchor=south,align=center] at (13.9,2.63) {$0$ queries\\[-1pt]{\footnotesize\citep{hadiji23towards}}};
\draw[lead] (14.75,2.68) -- (14.75,0.1);
}
{\hypersetup{citecolor=red!70!black}
\node[text=\lb,anchor=west,align=left] at (0.95,4.5) {$\Omega\bigl(\min\bigl\{d,\tfrac{\log(1/(d\varepsilon))}{\log d}\bigr\}\bigr)$\\[1.5pt]{\footnotesize\citep{hadiji23towards}}};
\draw[lead] (0.98,4.5) -- (0.24,4.5);
\node[text=\lb,anchor=west,align=left] at (0.95,2.65) {$\tilde\Omega\bigl(\min\{\varepsilon^{-2/3},\sqrt d\}\bigr)$\\[-1pt]{\footnotesize\citep{kornowski26oracle}}};
\node[text=\lb,anchor=north] at (5.0,1.65) {Corollary~\ref*{cor:query}};
}
\end{tikzpicture}
\caption{Upper bounds (blue solid) and lower bounds (red solid, dashed, and dotted) on the number of queries that deterministic algorithms need to output a pair $(\tilde x,\tilde y)$ with $\max_{x\in\Delta_d}\langle x,A\tilde y\rangle-\min_{y\in\Delta_d}\langle\tilde x,Ay\rangle\le2\varepsilon$ for every $A\in[-1,1]^{d\times d}$.
The figure is adapted from Figure~1 of~\citet{hadiji23towards}.
The red thick line is the lower bound of~\cref{cor:query}, and the shaded region shows where it is larger than the other lower bounds.
The value $c$ is an absolute constant.}
\label{fig:query}
\end{figure}
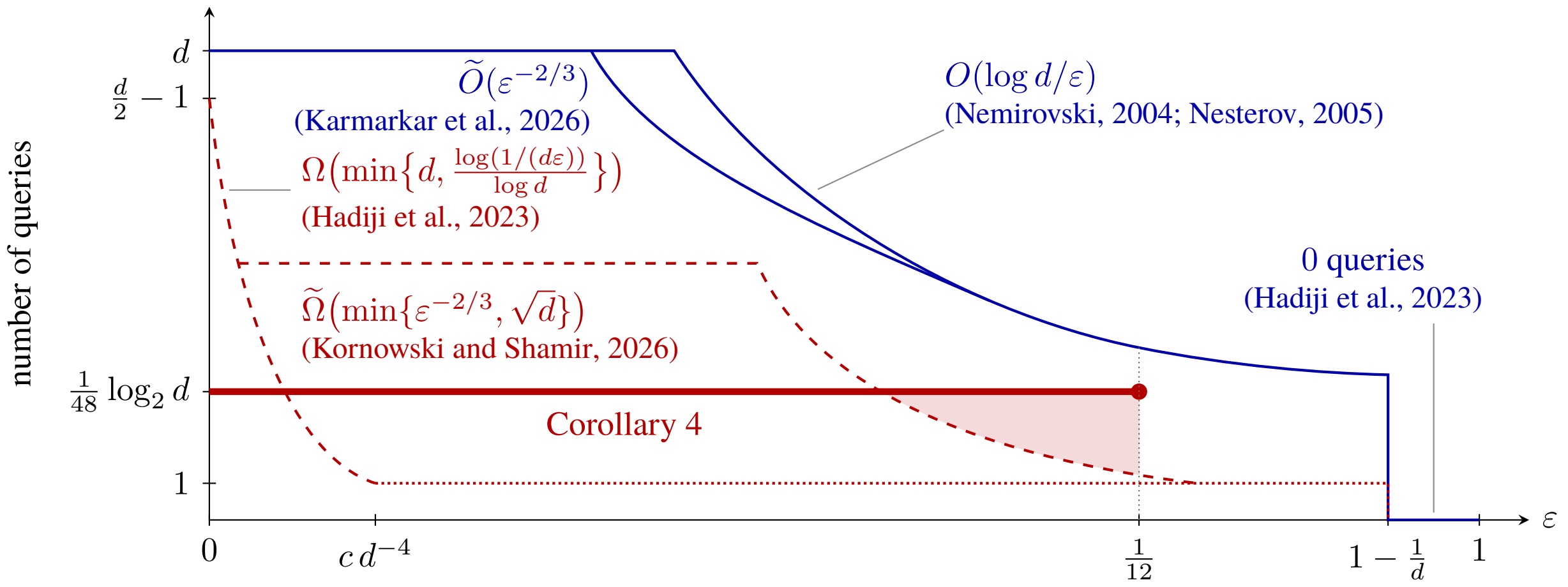

\paragraph{Query complexity of approximate Nash equilibria.}
In~\cref{sec:setting}, the goal of each player is to minimize its regret, which depends on the strategies selected in all rounds.
A different problem is to compute an approximate Nash equilibrium of an unknown payoff matrix $A\in[-1,1]^{m\times n}$ with a small number of queries.
In each query, an algorithm selects a pair of strategies $(x_t,y_t)\in\Delta_m\times\Delta_n$ and observes the products $Ay_t$ and $A^\top x_t$, which are the feedback vectors in~\cref{sec:setting}.
After the last query, the algorithm outputs a pair of strategies, and only this output is required to be an approximate Nash equilibrium.

The query complexity in this model is studied by~\citet{hadiji23towards} and \citet{karmarkar26solving}.
\citet{kornowski26oracle} study the model in which the queries may be arbitrary vectors.
The two models differ by at most a factor of two in the number of queries, since a product with an arbitrary vector is a linear combination of the products with two strategies.
In other settings, lower bounds are known for algorithms that observe the payoffs of pure strategy profiles~\citep{fearnley15learning} and for first-order methods for bilinear saddle-point problems over Euclidean balls~\citep{ouyang21lower}.
These lower bounds do not directly imply lower bounds in the query model above.

The best known upper bounds are $O(\sqrt{\log m\log n}/\varepsilon)$ queries~\citep{nemirovski04prox,nesterov05smooth} and $\tilde O(\varepsilon^{-2/3})$ queries~\citep{karmarkar26solving}, where $\tilde O$ hides polylogarithmic factors in $m$, $n$, and $1/\varepsilon$.
In addition, $\min\{m,n\}$ queries determine $A$ and hence suffice for every $\varepsilon$.
As for lower bounds, \citet{hadiji23towards} show that $\Omega(\min\{d,\log(1/(d\varepsilon))/\log d\})$ queries are necessary for $d\times d$ matrices when $\varepsilon$ is at most of order $d^{-4}$, and \citet{kornowski26oracle} show that deterministic algorithms need $\tilde\Omega(\varepsilon^{-2/3})$ queries when $m$ and $n$ are polynomially large in $1/\varepsilon$.
The lower bound of~\citet{kornowski26oracle} matches the upper bound of~\citet{karmarkar26solving} up to polylogarithmic factors, but neither this lower bound nor that of~\citet{hadiji23towards} increases with the numbers of actions when $\varepsilon$ is a constant.
\Cref{thm:basic} implies the following lower bound for deterministic algorithms, which increases with the numbers of actions.

\begin{corollary}\label{cor:query}
Let $m,n\ge4$ and $\varepsilon\in[0,1/6]$.
Consider a deterministic algorithm that, for every $A\in[-1,1]^{m\times n}$, sequentially selects $N$ pairs $(x_t,y_t)\in\Delta_m\times\Delta_n$, each depending on the previous pairs and on $(Ay_s,A^\top x_s)_{s<t}$, and then outputs $(\tilde x,\tilde y)\in\Delta_m\times\Delta_n$ satisfying
\[
\max_{x\in\Delta_m}\langle x,A\tilde y\rangle-\min_{y\in\Delta_n}\langle\tilde x,Ay\rangle\le\varepsilon.
\]
Then $N\ge\frac1{48}\log_2\min\{m,n\}$.
\end{corollary}

\begin{proof}
We prove the claim by applying~\cref{thm:basic} to the learning dynamic that selects the $N$ pairs of the algorithm and then its output.
Write $d=\min\{m,n\}$ and let $k=\lfloor\tfrac12\log_2d\rfloor$, which satisfies $k\ge1$ since $d\ge4$.
As in the proof of~\cref{cor:logarithmic}, we have $d\ge4^k\ge k2^k+1$ and $k\ge\tfrac14\log_2d$.
If $N\ge k$, then $N\ge\tfrac14\log_2d$, and the claim holds.
We therefore suppose that $N<k$.

Consider the following learning dynamic with $T=k$ rounds.
In each round $t\le N$, the two players select the $t$-th pair $(x_t,y_t)$ of the algorithm, and in each round $t>N$, they select its output $(\tilde x,\tilde y)$.
The feedback of the learning dynamic is $(g_s,\ell_s)=(Ay_s,A^\top x_s)$, and hence these strategies are functions of the history $(x_s,y_s,g_s,\ell_s)_{s=1}^{t-1}$.

By~\cref{thm:basic} with $T=k$, there exists a matrix $A\in[-1,1]^{m\times n}$ for which this learning dynamic satisfies $\Reg_k^x+\Reg_k^y\ge k/3$.
We next upper bound the social regret for this matrix.
For all $x^*\in\Delta_m$ and $y^*\in\Delta_n$, we have
\[
\langle x^*,Ay_t\rangle-\langle x_t,Ay^*\rangle\le
\begin{cases}
2&\text{for }t\le N,\\
\varepsilon&\text{for }t>N,
\end{cases}
\]
where the first case follows from $A\in[-1,1]^{m\times n}$, and the second from $(x_t,y_t)=(\tilde x,\tilde y)$ and the assumption on the output of the algorithm.
Hence we have
\[
\frac k3
\le\Reg_k^x+\Reg_k^y
=\max_{x^*\in\Delta_m,\;y^*\in\Delta_n}\sum_{t=1}^k\bigl(\langle x^*,Ay_t\rangle-\langle x_t,Ay^*\rangle\bigr)
\le2N+(k-N)\varepsilon
\le2N+\frac k6,
\]
where the first inequality follows from~\cref{thm:basic}, the equality from~\cref{eq:comparator}, and the last inequality from $\varepsilon\le1/6$.
Therefore $N\ge k/12\ge\frac1{48}\log_2d$, which completes the proof.
\end{proof}

An $\varepsilon$-approximate Nash equilibrium $(\tilde x,\tilde y)$ satisfies the inequality in~\cref{cor:query} with $2\varepsilon$ in place of $\varepsilon$, and hence the corollary applies to algorithms that output an $\varepsilon$-approximate Nash equilibrium with $\varepsilon\le1/12$.
For a constant $\varepsilon\in(0,1/12]$ and $m=n$, the lower bound matches the upper bound $O(\sqrt{\log m\log n}/\varepsilon)$ up to a constant factor.
The lower bound does not depend on $\varepsilon$, and hence it is much smaller than the upper bounds when $\varepsilon$ is small.
\Cref{fig:query} shows these bounds for $m=n=d$.

\paragraph{Open problem.}
Our lower bounds hold for deterministic learning dynamics, and it is an open problem whether they extend to learning dynamics that use internal randomization.
In~\cref{alg:construction}, the matrix $A$ is constructed from the strategies selected by the two algorithms.
If the algorithms use internal randomization, this matrix depends on the outcome of the randomization, and the construction does not give a fixed matrix for which the expected regrets satisfy the same lower bounds.
The lower bound of~\citet{kornowski26oracle} on the number of matrix-vector products is also restricted to deterministic algorithms.

\section*{AI Use Statement}
The final construction of the payoff matrices in our lower bounds was obtained using GPT-6 Astra.
The author substantially revised this construction and its presentation for clarity.
Claude Fable 5.1, 5.2, and Opus 5.5 were used to revise and organize the manuscript.
The author verified all proofs and is responsible for the content of this paper.

\section*{Acknowledgments}
TT is supported by JSPS KAKENHI Grant Number JP26K21297 and JST BOOST Grant Number JPMJBY25D4.

\bibliographystyle{plainnat}
\bibliography{references}

\newpage

\appendix

\section{Games with Two Columns}\label{app:two-columns}
\Cref{cor:rectangular} assumes that $\log m\le n^\alpha$ and that $n$ is sufficiently large.
The following proposition shows that these conditions cannot be removed entirely, since it gives a deterministic learning dynamic for $n=2$ whose regrets are at most $4$ for every $m$, whereas $\sqrt{\log m\log n}$ grows with $m$.

\begin{proposition}\label{prop:two-columns}
Let $T,m\ge1$ and $n=2$.
Then there exists a deterministic learning dynamic such that, for every $A\in[-1,1]^{m\times2}$ and every $H\in[T]$,
\[
\Reg_H^x\le4,
\qquad
\Reg_H^y\le4.
\]
\end{proposition}

\begin{proof}
In rounds $1$ and $2$, the players select $x_1=x_2=e_1$, $y_1=e_1$, and $y_2=e_2$.
The feedback of the $x$-player in these rounds consists of the two columns $g_1=Ae_1$ and $g_2=Ae_2$ of $A$, which both players can use at every round $t\ge3$, since each player may use the feedback of the opponent.
For every $t\ge3$, the players select $x_t=x^*$ and $y_t=y^*$, where $(x^*,y^*)$ is a Nash equilibrium of $A$ that both players compute from $g_1$ and $g_2$ by a fixed deterministic procedure.

For all $x\in\Delta_m$ and $y\in\Delta_2$, we have
\[
\langle x-x_t,Ay_t\rangle\le
\begin{cases}
2&\text{for }t\le2,\\
0&\text{for }t\ge3,
\end{cases}
\qquad
\langle y_t-y,A^\top x_t\rangle\le
\begin{cases}
2&\text{for }t\le2,\\
0&\text{for }t\ge3,
\end{cases}
\]
where the bounds for $t\le2$ follow from $A\in[-1,1]^{m\times2}$, and those for $t\ge3$ from $(x_t,y_t)=(x^*,y^*)$ and $\langle x,Ay^*\rangle\le\langle x^*,Ay^*\rangle\le\langle x^*,Ay\rangle$.
Hence, for every $H\in[T]$,
\[
\Reg_H^x
=\max_{x\in\Delta_m}\sum_{t=1}^H\langle x-x_t,Ay_t\rangle
\le4,
\qquad
\Reg_H^y
=\max_{y\in\Delta_2}\sum_{t=1}^H\langle y_t-y,A^\top x_t\rangle
\le4,
\]
which completes the proof.
\end{proof}

\end{document}